\documentclass[aps,preprint,floatfix]{revtex4}
\usepackage{epsfig}
\usepackage{amsmath,amsfonts,bm}
\begin{document}

\count255=\time\divide\count255 by 60 \xdef\hourmin{\number\count255}
  \multiply\count255 by-60\advance\count255 by\time
 \xdef\hourmin{\hourmin:\ifnum\count255<10 0\fi\the\count255}

\newcommand\<{\langle}
\renewcommand\>{\rangle}
\renewcommand\d{\partial}
\newcommand\LambdaQCD{\Lambda_{\textrm{QCD}}}
\newcommand\tr{\mathop{\mathrm{Tr}}}
\newcommand\+{\dagger}
\newcommand\g{g_5}

\newcommand{\xbf}[1]{\mbox{\boldmath $ #1 $}}

\title{Minimal Lee-Wick Extension of the Standard Model}

\author{Christopher D. Carone}
\email{cdcaro@wm.edu}

\affiliation{Department of Physics, College of William \& Mary,
Williamsburg, VA 23187-8795}

\author{Richard F. Lebed}
\email{Richard.Lebed@asu.edu}

\affiliation{Department of Physics, Arizona State University, Tempe,
AZ 85287-1504}

\date{June 2008}

\begin{abstract}
We consider a minimal Lee-Wick (LW) extension to the Standard Model in
which the fields providing the most important contributions to the
cancellation of quadratic divergences are the lightest.  Partners to
the SU(2) gauge bosons, Higgs, top quark, and left-handed bottom quark
are retained in the low-energy effective theory, which is valid up to
approximately $10$~TeV; the remaining LW partners appear above this
cutoff and complete the theory in the ultraviolet.  We determine the
constraints on the low-energy spectrum from the electroweak parameters
$S$ and $T$, and find LW states within the kinematic reach of the LHC
at the 95\% confidence level.
\end{abstract}


\maketitle

\section{Introduction} \label{intro}

The intriguing idea of Lee and Wick (LW)~\cite{Lee:1969fy} to promote
Pauli-Villars regulators to the status of physical fields was recently
applied to develop a LW extension to the Standard Model
(LWSM)~\cite{Grinstein:2007mp}.  While the original LW proposal was
designed to render QED finite, the purpose of the LWSM is to use the
LW opposite-sign propagators in loop diagrams to solve the hierarchy
problem.  This solution is analogous to the supersymmetric one in that
it relies on cancellation between pairs of loops to remove quadratic
divergences, but differs in that the LW particles carry the same
statistics (and other quantum numbers) as their SM partners.  Several
recent papers investigate the formal properties and phenomenology of
the LWSM~\cite{lwpapers,Dulaney:2007dx,Krauss:2007bz,Alvarez:2008za,Underwood:2008cr}.

In the present work, we consider a version of the LWSM in which only a
subset of the full spectrum of LW partners lie within the reach of the
LHC.  Motivated by Little Higgs models~\cite{lhm}, we study the
possibility that only the LW partners of the SU(2) gauge bosons,
Higgs, $t$ quark, and left-handed $b$ quark appear in the low-energy
effective theory.  These fields provide the most significant
contributions to the cancellation of quadratic divergences in the
Higgs sector and render the effective theory natural, provided the
cutoff is $\alt 10$~TeV.  The remaining LW spectrum may appear above
this cutoff, or the theory may be completed by other, more exotic
physics.  This minimal LW low-energy theory is distinguished by its
simplicity, making it an ideal subject for comprehensive
phenomenological investigation.

In this Letter we present the constraints on this model's spectrum
that follow from oblique electroweak parameters, in particular the
Peskin-Takeuchi $S$ and $T$ parameters~\cite{PandT}:
\begin{eqnarray}
S & = & -16 \pi \frac{d}{dq^2}\Pi_{3B} |_{q^2=0} \,,  \label{eq:sdef}
\\
T & = & \frac{4 \pi}{s^2 c^2 m_{Z_0^2}} \left(
\Pi_{11}-\Pi_{33}\right) |_{q^2=0} \, ,\label{eq:tdef}
\end{eqnarray}
where $\Pi$ are the usual self-energy functions, $s \! \equiv \!
\sin \theta_W$, $c \! \equiv \! \cos \theta_W$ parametrize the weak
mixing angle and $m_{Z_0}$ is the measured $Z$ boson mass.  Our
approach is similar to that of other recent work, in particular
Ref.~\cite{Alvarez:2008za} (ADSS), but differs in that we do not
assume a complete LWSM spectrum with large sets of mass-degenerate
particles.  Exact one-loop formulae for $S$ and $T$, which have not
appeared in previous literature, are necessary for a proper treatment
of corrections in our model. We agree with the original LWSM
work~\cite{Grinstein:2007mp} and ADSS~\cite{Alvarez:2008za} that the
leading oblique corrections occur at tree level in the LWSM, contrary
to the claim in Ref.~\cite{Underwood:2008cr}.  Part of this discrepancy 
is due to differing definitions in the literature of what physics is ``oblique'' (see 
the discussion in Sec.~\ref{conv}).  Moreover, the identification of oblique parameters 
in the effective Lagrangian of Ref.~\cite{Underwood:2008cr} appears to be fundamentally 
different from ours and Refs~\cite{Grinstein:2007mp,Alvarez:2008za} so that the numerical 
results are not easily compared; we do not consider this issue further here.

Our paper is organized as follows: In Sec.~\ref{conv} we establish
conventions for specifying the spectrum, which take into account
potentially substantial mixing between SM and LW particles.  We
present the one-loop formulae for the $S$ and $T$ in
Sec.~\ref{sec:loops}.  We present our numerical results in
Sec.~\ref{sec:results}, and Sec.~\ref{sec:concl} summarizes our
conclusions.

\section{Preliminaries} \label{conv}

We study the $S$ and $T$ parameters in an effective theory obtained by
integrating out heavy-mass eigenstates.  At tree level, this procedure
is equivalent to eliminating the heavy fields from the Lagrangian
using their classical equations of motion.  Since the gauge sector of
our model includes LW partners to only the SU(2) gauge bosons, one
finds
\begin{eqnarray} \label{eq:stree}
\Delta S_{\rm tree} & = & 4\pi \frac{v^2}{M_2^2} + O \left(
\frac{v^4}{M_2^4} \right) \, , \\ \Delta T_{\rm tree} & = & 0 \,,
\label{eq:ttree}
\end{eqnarray}
in agreement with the results of ADSS in the limit $M_1 \! \to \!
\infty$ [The U(1) LW gauge boson contributes to $\Delta T_{\rm tree}$
at $O (v^2/M_1^2)$]. Here, $M_1$ and $M_2$ represent the unmixed LW
U(1) and SU(2) gauge boson masses in the auxiliary field formulation
of the LWSM, as defined in Ref.~\cite{Grinstein:2007mp}, and $v$ is
the Higgs vacuum expectation value.  The parameter $U$ turns out to be
$O(v^4 / M_2^4)$ and is similarly suppressed in loop effects, so we do
not consider it further.  Note that the constraints on new physics
from oblique parameters are meaningful only if vertex corrections are
small.  The derivation of Eqs.~(\ref{eq:stree})--(\ref{eq:ttree})
includes field redefinitions that force the couplings of the gauge
fields to SM currents to match those of the SM at tree level.  Thus,
the definitions of $S$ and $T$ used here (and in ADSS) subsume the
largest vertex corrections.

\begin{figure}
\includegraphics[scale=.8]{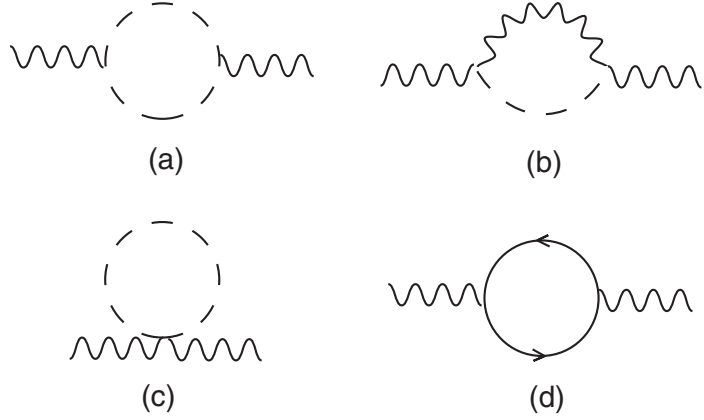}
\caption{\label{fig:one} Diagram classes that may contribute to
oblique parameters.  Wavy lines represent gauge fields, dashed lines
represent scalars, and solid lines represent fermions.}
\end{figure}

The one-loop contributions to the self-energies $\Pi_{AB}$ in
Eqs.~(\ref{eq:sdef})--(\ref{eq:tdef}) arise from the diagrams in
Fig.~\ref{fig:one}.  We evaluate these diagrams using mass eigenstates
on the internal lines; since SM and LW fields mix, one must first
define conventions to specify the spectrum.  The following mixing
effects are taken into account:

\noindent 1. {\em Neutral Higgs mixing.}  The SM Higgs field and its
LW partner $(h , \tilde{h})$ have mass terms~\cite{Grinstein:2007mp}
\begin{equation}\label{eq:hhtmm}
\delta {\cal L} = -\frac 1 2 \left( \begin{array}{cc} h & \tilde h
\end{array} \right) \left(\begin{array}{cc} m_h^2 & -m_h^2 \\ -m_h^2 &
-(m_{\tilde{h}}^2 - m_h^2) \end{array}\right) \left( \begin{array}{c}
h \\ \tilde h \end{array} \right) \, .
\end{equation}
The mass matrix in Eq.~(\ref{eq:hhtmm}) is diagonalized via the
symplectic transformation
\begin{equation} \label{symprot}
\left( \begin{array}{c} h \\ \tilde h \end{array} \right) = \left(
\begin{array}{cc} \cosh \theta & \sinh \theta \\ \sinh \theta & \cosh
\theta \end{array} \right) \left( \begin{array}{c} h_0 \\ \tilde h_0
\end{array} \right) \, ,
\end{equation}
where subscript $0$ here and below indicates mass eigenstates.  The
mixing angle $\theta$ satisfies
\begin{equation} \label{scalar_eigen}
\tanh 2\theta = \frac{-2m_h^2/m_{\tilde h}^2}{1-2m_h^2/m_{\tilde h}^2}
= -\frac{2m_{h_0}^2 m_{\tilde h_0}^2}{m_{h_0}^4 +
m_{\tilde h_0}^4} \, ,
\end{equation}
with mass eigenvalues
\begin{equation}
m_{h_0}^2, \, m_{\tilde h_0}^2 = \frac{m_{\tilde h}^2}{2} \left( 1 \mp
\sqrt{ 1 -\frac{4 m_h^2}{m_{\tilde h}^2} } \right) \, .
\end{equation}
In addition, the LW sector has pseudoscalar $\tilde P$ and charged
scalar $\tilde h^+$ states with masses $m_{\tilde h}$. We
work in unitary gauge, where all unphysical scalars are eliminated
from the theory.

\noindent 2. {\em Gauge mixing.}
The SM SU(2) gauge boson and its LW partner ($W \, , \,
\widetilde{W}$) mix via~\cite{Grinstein:2007mp}
\begin{equation}
\delta {\cal L} = \left( \begin{array}{cc} W^{\mu +} & \widetilde
W^{\mu +} \end{array} \right)
\left( \begin{array}{cc} m_W^2 & m_W^2 \\ m_W^2 &
m_W^2 \! - \! M_2^2 \end{array} \right) \left( \begin{array}{c}
W^-_\mu \\ \widetilde W^-_\mu \end{array} \right) \, ,
\end{equation}
where $m_W \! = \! \frac 1 2 g_2 v$ is the unmixed SM $W$ mass.  The
mass matrix is diagonalized by the symplectic transformation
\begin{equation}
\left( \begin{array}{c} W^\pm \\ \widetilde W^\pm \end{array} \right)
= \left( \begin{array}{cc} \cosh \varphi_c & \sinh \varphi_c \\ \sinh
\varphi_c & \cosh \varphi_c \end{array} \right) \left(
\begin{array}{c} W^\pm_0 \\ \widetilde W^\pm_0 \end{array} \right) \,
,
\end{equation}
where, using $\widetilde W_0^1$ to indicate the charged heavy mass
eigenstate,
\begin{equation} \label{vectorceigen}
\tanh 2\varphi_c = \frac{2m_W^2}{M_2^2 - 2m_W^2} = \frac{2 m_{W_0}^2
m_{\widetilde W_0^1}^2}{m_{W_0}^4 + m_{\widetilde W_0^1}^4} \, ,
\end{equation}
with eigenvalues satisfying
\begin{equation}
m_{W_0}^2, \ m_{\widetilde W_0^1}^2 = \frac{M_2^2}{2} \left( 1 \mp
\sqrt{ 1 - \frac{4m_W^2}{M_2^2}} \right) \, .
\end{equation}
In the neutral sector, mixing only occurs between the SM $Z$ boson
and the LW $\widetilde{W}^3$~\cite{Grinstein:2007mp}:
\begin{equation}
\delta {\cal L} = \frac 1 2 \left( \begin{array}{cc} Z &
\widetilde{W}^3 \end{array} \right) \left(\begin{array}{cc} m_Z^2 &
m_Z^2 \,c \\ m_Z^2 \,c & -(M_2^2-m_W^2) \end{array}\right) \left(
\begin{array}{c} Z \\
\widetilde{W}^3 \end{array} \right) \, ,
\label{eq:zw3mm}
\end{equation}
where $m_Z=m_W/c$ is the unmixed SM $Z$ mass. The photon decouples as
a consequence of electromagnetic gauge invariance.
Equation~(\ref{eq:zw3mm}) is diagonalized via the symplectic
transformation
\begin{equation}
\left( \begin{array}{c} Z \\ \widetilde W^3 \end{array} \right) =
\left( \begin{array}{cc} \cosh \varphi_0 & \sinh \varphi_0 \\ \sinh
\varphi_0 & \cosh \varphi_0 \end{array} \right) \left(
\begin{array}{c} Z_0 \\ \widetilde W^3_0 \end{array} \right) \, ,
\end{equation}
where
\begin{equation} \label{vector0eigen}
\tanh 2\varphi_0 = \frac{2m_Z^2 c}{M_2^2 - m_Z^2 (1+c^2)} \, ,
\end{equation}
and the eigenvalues are given by
\begin{equation}
m_{Z_0}^2, \, m_{\widetilde W_0^3}^2 = \frac 1 2 \left[ M_2^2 + m_Z^2
s^2 \mp \sqrt{ (M_2^2 - m_Z^2 s^2)^2 - 4M_2^2 m_Z^2 c^2} \right] \, .
\end{equation}
\noindent 2. {\em Fermion mixing.}
Our model includes LW partners to the fields $t_L$, $t_R$, and $b_L$.
The mass terms of the third-generation fermions
read~\cite{Krauss:2007bz,Alvarez:2008za}
\begin{equation}
\delta {\cal L} = - \overline{T}^{\vphantom\dagger}_L \eta
{\cal M}_t^\dagger T^{\vphantom\dagger}_R -
\overline{B}^{\vphantom\dagger}_L \eta
{\cal M}_b^\dagger B^{\vphantom\dagger}_R + {\rm h.c.} \, ,
\end{equation}
where
\begin{eqnarray}
T_{L, R}^T & = & ( t_{L, R}^{\vphantom\prime}, \, \tilde t_{L,
R}^{\vphantom\prime}, \, \tilde t_{L, R}^{\; \prime} ) \, ,
\label{tmult} \\
B_{L, R}^T & = & ( b_{L, R}^{\vphantom\prime}, \, \tilde b_{L,
R}^{\vphantom\prime}, \, \tilde b_{L, R}^{\; \prime} ) \, ,
\label{bmult}
\end{eqnarray}
define our basis for the third-generation fields.  The fields $t_L$,
$t_R$, $b_L$, and $b_R$ are the SM fields with their usual quantum
numbers, while the tilded fields are LW.  The unprimed LW fields are
the partners of the SM fields and hence have the same quantum numbers
and chirality.  The primed LW fields have the same quantum numbers as
the unprimed LW fields of the opposite chirality, in order to permit
SU(2)$\times$U(1)-invariant LW mass terms.  Thus, for example,
$\tilde t_L$ and $\tilde t_R^{\; \prime}$ both transform as a (${\bf
2}$, $+\frac 1 6$) under SU(2)$\times$U(1), the same as $t_L$.  The
matrix $\eta = \rm{diag} ( 1, \, -1, \, -1 )$ conveniently encodes the
opposite signs between SM and LW kinetic terms or mass terms.  Then
one finds
\begin{equation} \label{mmatrices}
{\cal M}_t \eta = \left( \begin{array}{ccc} +m_t & -m_t & 0 \\ -m_t &
+m_t & -M_t \\ 0 & -M_q & 0 \end{array} \right) \, , \
{\cal M}_b \eta = \left( \begin{array}{ccc} +m_b & -m_b & 0 \\ -m_b &
+m_b & -M_b \\ 0 & -M_q & 0 \end{array} \right) \, .
\end{equation}
We diagonalize these mass matrices via transformation matrices $S^a_L$
and $S^a_R$, for $a=t$ or $b$, such that ${\cal M}_0$ is diagonal with
positive eigenvalues:
\begin{equation} \label{massxfm}
S^\dagger_L \eta S_L = \eta \, , \ S^\dagger_R \eta S_R = \eta \, , \
{\cal M}_0 \eta = S^\dagger_R {\cal M} \eta S_L \, .
\end{equation}
Additional details regarding the solution to Eqs.~(\ref{massxfm}) will
appear elsewhere~\cite{clnext}; for the purposes of this calculation,
we simply note that solutions were obtained numerically.

\section{Loops}\label{sec:loops}

A consistent calculation of oblique parameters in a perturbative
theory must yield results that are ultraviolet finite, since these
parameters describe physical observables. Here we consider the
deviation of $S$ and $T$ from their SM values, so one must subtract
any purely SM contributions.  While individual diagrams can diverge,
we find that the final subtracted results are finite and cutoff
independent.

First consider the $S$ parameter, which receives contributions from
the diagrams of Fig.~\ref{fig:one}a, b, and d; the diagram in
Fig.~\ref{fig:one}c is not relevant since its contributions to
$\Pi_{3B}$ is $q^2$ independent. From the purely Higgs-sector diagram
in Fig.~\ref{fig:one}a we find
\begin{equation} \label{S1a}
\Delta S_{1\rm a} = \frac{1}{12\pi} \left[ I_1 ( m^2_{\tilde
h_0}/m^2_{\tilde h} ) \cosh^2 \theta - I_1 ( m^2_{h_0}/m^2_{\tilde h}
) \sinh^2 \theta \right] \, ,
\end{equation}
where
\begin{equation} \label{I1def}
I_1 (\xi) \equiv \frac{\xi^2 (3 - \xi) \ln \xi}{(1-
\xi)^3} - \frac{(5 - 22\xi + 5\xi^2)}{6 (1 - \xi)^2} \, .
\end{equation}
Note that the contribution to the self-energy from Fig.~\ref{fig:one}a
vanishes if the LW states are decoupled, so the result must be finite
without any SM subtraction, as is indeed the case.  The contribution
from Fig.~\ref{fig:one}b, however, involves a diagram with purely SM
particles (the Higgs and Z bosons), with non-SM couplings.  In this
case, one must subtract the same diagrams evaluated with infinite LW
masses.  One finds
\begin{eqnarray}
\Delta S_{1\rm b} & = & -\frac{g_2^2 v^2}{4\pi M_2^2} \sum
\frac{C_{3B}}{\xi_2} \left[ I_2 \left( \frac{\xi_1}{\xi_2} \right) -
\frac{1}{12} I_1 \left( \frac{\xi_1}{\xi_2} \right) - \frac{1}{12} \ln
\xi_2 \right] \, , \label{S1b}
\end{eqnarray}
where $C_{3B}$ is the coefficient of a Fig.\ref{fig:one}b diagram with
internal scalar (S) and vector (V) particles of mass $m_S$ and $m_V$,
respectively, $\xi_1 \equiv m_S^2/M_2^2$, $\xi_2 \equiv m_V^2/M_2^2$,
and
\begin{equation} \label{I2def}
I_2 (\xi) \equiv \frac{1 - \xi^2 + 2 \xi \ln \xi}{2 (1 - \xi)^3} \, .
\end{equation}
Table~\ref{Table1b} gives the values of $C_{3B}$, $\xi_1$, and $\xi_2$
for each term summed in Eq.~(\ref{S1b}), as well as for the SM
subtraction.

\begin{table}[t]\vspace{-6pt}
\caption[]{Coefficients $C_{AB}$ for each of the contributing diagrams
to Eq.~(\ref{S1b}) and (\ref{T1b}).}
\begin{ruledtabular}
\begin{tabular}{cccccc}
$AB$ & $C_{AB}$ & $S$ & $V$ & $\xi_1$ & $\xi_2$ \\
\hline
$3B$ & $+\cosh^2 \theta \left( \sinh \varphi_0 + \frac{1}{c} \cosh
\varphi_0 \right)^2$ & $h_0$ & $Z_0$ & $m_{h_0}^2/M_2^2$ &
$m_{Z_0}^2/M_2^2$ \\
     & $-\cosh^2 \theta \left( \cosh \varphi_0 + \frac{1}{c} \sinh
\varphi_0 \right)^2$ & $h_0$ & $\widetilde W^3_0$ & $m_{h_0}^2/M_2^2$
& $m_{\widetilde W^3_0}^2/M_2^2$ \\
     & $-\sinh^2 \theta \left( \sinh \varphi_0 + \frac{1}{c} \cosh
\varphi_0 \right)^2$ & $\tilde h_0$ &          $Z_0$ &
$m_{\tilde h_0}^2/M_2^2$ &          $m_{Z_0}^2/M_2^2$ \\
     & $+\sinh^2 \theta \left( \cosh \varphi_0 + \frac{1}{c} \sinh
\varphi_0 \right)^2$ & $\tilde h_0$ & $\widetilde W^3_0$ &
$m_{\tilde h_0}^2/M_2^2$ & $m_{\widetilde W^3_0}^2/M_2^2$ \\
     &  $-\frac{1}{c^2}$ &      $h$ &            $Z$ &
           $m_h^2/M_2^2$ &              $m_Z^2/M_2^2$ \\
\hline
$11$ & $-\cosh^2 \theta (\cosh \varphi_c + \sinh \varphi_c)^2$ & $h_0$
& $W^1_0$        & $m_{h_0}^2/M_2^2$ & $m_{       W^1_0}^2/M_2^2$ \\
     & $+\cosh^2 \theta (\cosh \varphi_c + \sinh \varphi_c)^2$ & $h_0$
& $\widetilde W^1_0$ & $m_{h_0}^2/M_2^2$ &
$m_{\widetilde W^1_0}^2/M_2^2$ \\
     & $+\sinh^2 \theta (\cosh \varphi_c + \sinh \varphi_c)^2$ &
$\tilde h_0$ & $       W^1_0$ & $m_{\tilde h_0}^2/M_2^2$ &
$m_{       W^1_0}^2/M_2^2$ \\
     & $-\sinh^2 \theta (\cosh \varphi_c + \sinh \varphi_c)^2$ &
$\tilde h_0$ & $\widetilde W^1_0$ & $m_{\tilde h_0}^2/M_2^2$ &
$m_{\widetilde W^1_0}^2/M_2^2$ \\
     &                                                    $+1$ &
         $h$ &          $W^1$ &            $m_h^2/M_2^2$ &
         $m_{W^1}^2/M_2^2$ \\
\hline
$33$ & $-\cosh^2 \theta \left( \sinh \varphi_0 + \frac{1}{c} \cosh
\varphi_0 \right)^2$ &        $h_0$ &          $Z_0$ &
$m_{h_0}^2/M_2^2$        &          $m_{Z_0}^2/M_2^2$ \\
     & $+\cosh^2 \theta \left( \cosh \varphi_0 + \frac{1}{c} \sinh
\varphi_0 \right)^2$ &        $h_0$ & $\widetilde W^3_0$ &
$m_{h_0}^2/M_2^2$        & $m_{\widetilde W^3_0}^2/M_2^2$ \\
     & $+\sinh^2 \theta \left( \sinh \varphi_0 + \frac{1}{c} \cosh
\varphi_0 \right)^2$ & $\tilde h_0$ &          $Z_0$ &
$m_{\tilde h_0}^2/M_2^2$ &          $m_{Z_0}^2/M_2^2$ \\
     & $-\sinh^2 \theta \left( \cosh \varphi_0 + \frac{1}{c} \sinh
\varphi_0 \right)^2$ & $\tilde h_0$ & $\widetilde W^3_0$ &
$m_{\tilde h_0}^2/M_2^2$ & $m_{\widetilde W^3_0}^2/M_2^2$ \\
     &  $+\frac{1}{c^2}$ &      $h$ &            $Z$ &
           $m_h^2/M_2^2$ &              $m_Z^2/M_2^2$ \\
\end{tabular}
\end{ruledtabular}
\label{Table1b}
\end{table}
To compute the fermionic contribution to $S$, we first parametrize the
gauge-fermion couplings evaluated in the mass eigenstate basis:
\begin{equation}\label{eq:candd}
\delta {\cal L} = -g_1 B_\mu \overline{\Psi}_0 \gamma^\mu ( C^{L}_\Psi
P_L + C^R_\Psi P_R ) \Psi_0 -g_2 W^3_\mu \overline{\Psi}_0 \gamma^\mu
( D^L_\Psi P_L + D^R_\Psi P_R ) \Psi_0 \, ,
\end{equation}
where $P_L$ ($P_R$) are the left (right)-handed chiral projection
operators, and $\Psi_0$ represents $T_0$ and $B_0$, the transformation
of Eq.~(\ref{tmult}) and (\ref{bmult}), respectively, into mass
eigenstates.  The gauge coupling matrices $C^{L,R}_\Psi$ and
$D^{L,R}_\Psi$ are computed numerically, taking into account the basis
change Eq.~(\ref{massxfm}).  Denoting the mass of the $i^{\rm th}$
fermion mass eigenstate $m_i$, and defining $\xi_i \! \equiv \!
m_i^2/M^2$ for an arbitrary mass scale $M$, we find
\begin{eqnarray}
\Delta S_{1\rm d} & = & -\frac{2}{\pi} \sum_{\Psi=T,B} \sum_{i,j}
\eta^{\vphantom\dagger}_{ii} \eta^{\vphantom\dagger}_{jj}
\left\{ ( C_{\Psi \, ij}^L D_{\Psi \,j \, i}^L + C_{\Psi \, ij}^R
D_{\Psi \,j \, i}^R ) \left[ I_1
\left( \frac{\xi_i}{\xi_j} \right) + \ln \xi_j \right]
\right. \nonumber \\ & & \left.
- 3 ( C_{\Psi \, ij}^L D_{\Psi \,j \, i}^R + C_{\Psi \, ij}^R
D_{\Psi \,j \, i}^L ) \sqrt{
\frac{\xi_i}{\xi_j}} I_2 \left( \frac{\xi_i}{\xi_j} \right)
\right\} - \frac{1}{2\pi}\left[1-\frac{1}{3}\ln
\left(\frac{m^2_{t, \, \rm SM}}{m^2_{b,\, \rm SM}}\right)\right]\,.
\label{Sfermion}
\end{eqnarray}
The last term of Eq.~(\ref{Sfermion}) represents the SM subtraction,
with $m_{t, \, \rm SM}$ and $m_{b, \, \rm SM}$ the $t$ and $b$ masses,
respectively, obtained in the decoupling limit of the LW states.  The
cancellation of logarithmic divergences between various contribution
to $\Delta S_{1\rm d}$ requires
\begin{equation}
\sum_{i,j} \eta^{\vphantom\dagger}_{ii} \eta^{\vphantom\dagger}_{jj}
( C_{ij}^L D_{j \, i}^L + C_{ij}^R D_{j \, i}^R ) = 0 \, ,
\end{equation}
which we find to be satisfied to any desired numerical precision.  The
numerical results for $S$ presented in the next section represent the
total $\Delta S \! = \! \Delta S_{1\rm a} \! + \! \Delta S_{1\rm b} \!
+ \! \Delta S_{1\rm d}$ given by Eqs.~(\ref{S1a}), (\ref{S1b}) and
(\ref{Sfermion}).

Our approach to evaluating $T$ is analogous.  In agreement with ADSS,
we find that the contributions to $T$ from Fig.~\ref{fig:one}a exactly
cancel, as do those from Fig.~\ref{fig:one}c.  The coefficients
$C_{11}$, $C_{33}$ for the diagrams in Fig.~\ref{fig:one}b, including
SM subtractions, appear in Table~\ref{Table1b}.  We find
\begin{eqnarray}
\Delta T_{1\rm b}  &=& -\frac{m_Z^2}{4\pi s^2 m_{Z_0}^2} \sum
\nonumber \\
&&\left(\frac{C_{11}}{\xi_1 - \xi_2} \left[ \xi_1 \ln \xi_1 - \xi_2
\ln \xi_2 - \frac{1}{4\xi_2} \left( \xi_1^2 \ln \xi_1 - \xi_2^2 \ln
\xi_2
\right) \right] - \left(C_{11} \to C_{33}\right) \right) \, ,
\label{T1b}
\end{eqnarray}
with $\xi_1$ and $\xi_2$ defined after Eq.~(\ref{S1b}).  To find the
fermionic contribution to $T$, we extend the parametrization of
gauge-fermion couplings of Eq.~(\ref{eq:candd}) to include the $W^1$
boson:
\begin{equation}
\delta {\cal L} = -g_2 W^1_\mu \overline{T}_0 \gamma^\mu (E_L P_L +
E_R P_R) B_0 + \mbox{ h.c.} \, ,
\end{equation}
where the matrices $E_{L,R}$ are also evaluated in the mass
eigenstate basis.  One finds
\begin{eqnarray}
\Delta T_{1\rm d} &=& -\frac{3}{4 \pi s^2 c^2 m_{Z_0}^2} \left\{
M^2 \! \sum_{\Psi=T,B} \sum_{i,j}\right. \eta^{\vphantom\dagger}_{ii}
\eta^{\vphantom\dagger}_{jj} \left[ -( D_{\Psi\,ij}^L
D_{\Psi\, j \, i}^L \! + D_{\Psi\, ij}^R D_{\Psi\,j \,
i}^R ) \left(\frac{\xi_i^2 \ln \xi_i - \xi_j^2 \ln \xi_j}{\xi_i -
\xi_j}\right) \right. \nonumber \\ &+& \left. 4\, ( D_{\Psi\,ij}^L
D_{\Psi\,j \, i}^R + D_{\Psi\,ij}^R D_{\Psi\,j \, i}^L )
\sqrt{\xi_i \xi_j} \left(\frac{\xi_i \ln \xi_i - \xi_j \ln \xi_j}
{\xi_i - \xi_j} \right)\right] \nonumber \\
&+& M^2 \sum_{i,j}
\eta^{\vphantom\dagger}_{ii} \eta^{\vphantom\dagger}_{jj} \left[
2\,( E_{ij}^L E_{j
\, i}^{L \, \dagger} + E_{ij}^R E_{j \, i}^{R \, \dagger} )
\left(\frac{\xi_i^2 \ln \xi_i - \xi_j^2 \ln \xi_j}{\xi_i - \xi_j}
\right) \right. \nonumber \\ &-& \left.  8 \,( E_{ij}^L E_{j
\, i}^{R \, \dagger} + E_{ij}^R E_{j \, i}^{L \, \dagger} )
\sqrt{\xi_i \xi_j} \left(\frac{\xi_i \ln \xi_i - \xi_j \ln \xi_j}
{\xi_i - \xi_j} \right)\right] \nonumber \\
&-&\left.\frac{1}{4} \left[m_{t, \, \rm SM}^2+m_{b, \, \rm SM}^2 -
\frac{ 2 m_{t, \, \rm SM}^2 m_{b,\, \rm SM}^2}{(m_{t, \, \rm
SM}^2-m_{b, \, \rm SM}^2)} \ln\left(\frac{m_{t, \, \rm SM}^2}{m_{b, \,
\rm SM}^2} \right)\right]\right\}\,. \label{T1d}
\end{eqnarray}
The removal of divergences from $\Delta T_{1\rm d}$ [leading to the
finiteness of Eq.~(\ref{T1d})] requires delicate cancellations between
the $t$, $b$, and $tb$ diagrams not only for the $LL \! + \! RR$
coefficients of the quadratic divergences, but also between the $LL \!
+ \! RR$ and $LR \! + \! RL$ coefficients of the logarithmic
divergences.  Indeed, these cancellations may be verified~\cite{clnext}.

\section{Results}\label{sec:results}

To obtain the constraints on our model, we choose as input parameters
$M_2$, $m_{\tilde{h}}$, and a common fermion mass parameter $M_F \! =
\! M_q \! = \! M_t$.  With the lightest gauge boson mass eigenvalues
$m_{W_0}$, $m_{Z_0}$ fixed by the measured masses, specifying $M_2$
fixes the gauge boson spectrum of the model.  We choose the Higgs mass
parameter $m_h$ that appears in the SM Lagrangian to be $115$~GeV,
which provides our reference mass in defining $S$ and $T$; specifying
$m_{\tilde{h}}$ then completely fixes the Higgs spectrum of the
theory.  Finally, we set the lightest fermion mass eigenvalues
$m_{t_0}$, $m_{b_0}$ to the physical quark masses, and decouple the LW
partner $\tilde b_R$, which is not part of the minimal low-energy
theory, by taking $M_b \rightarrow \infty$; specifying $M_F$ then
completely fixes the fermionic spectrum of the theory.  Note that the
choice $M_F \! = \! M_q \! = \! M_t$ is merely a convenience, although
naturalness suggests $M_q$ and $M_t$ are comparable; the general case
provides substantial additional freedom in accommodating current
experimental bounds~\cite{clnext}.  Finally, note that each set of
input parameters $(M_2,m_{\tilde{h}},M_F)$ corresponds to a (slightly)
different value for the Lagrangian mass parameter $m_t$.  This mass
corresponds to the SM reference value in the limit of decoupled LW
partners, about which deviations in $S$ and $T$ are measured.  We
shift~\cite{PandT} the model predictions for the oblique parameters
for each input parameter set to coincide with the $t$ reference mass
170.9~GeV assumed in the computation of the experimentally allowed
region of the $S$-$T$ plane.

\begin{figure}
\includegraphics[scale=.75]{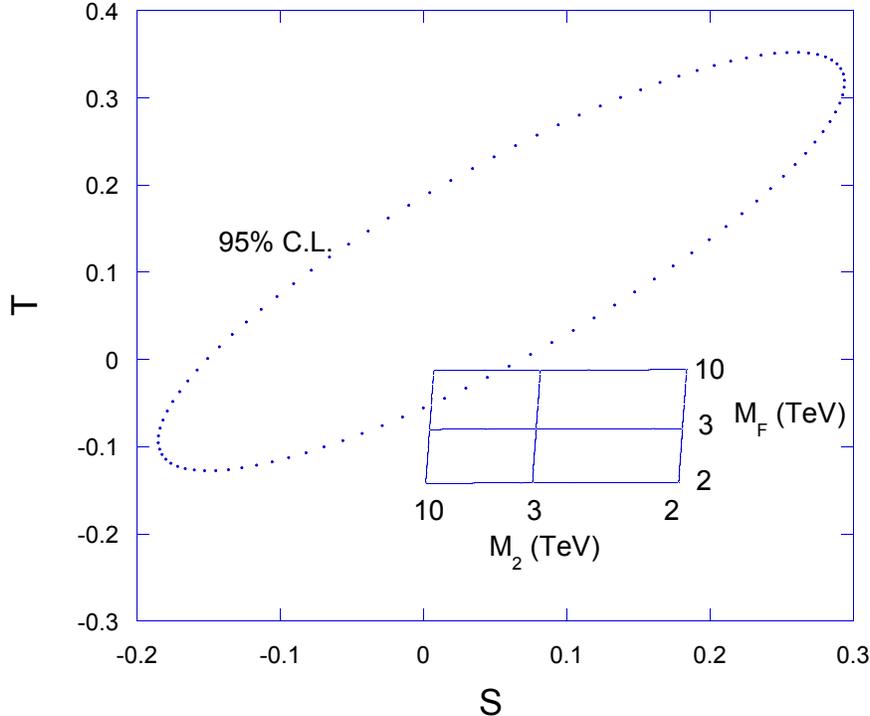}
\caption{\label{fig:two} Oblique corrections for
$m_{\tilde{h}}=750$~GeV.  The grid shows model predictions as
$M_2$ and $M_F$ are varied from 2--10~TeV. The Higgs and $t$ reference 
masses are $115$~GeV and $170.9$~GeV, respectively.}
\end{figure}

\begin{figure}
\includegraphics[scale=.75]{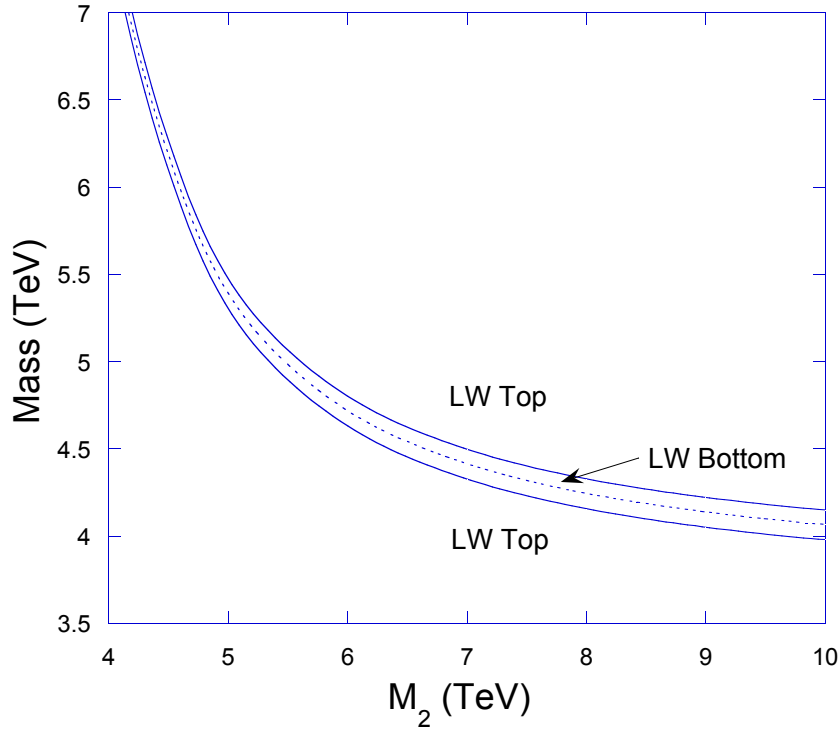}
\caption{\label{fig:three} Limits on the smallest 95\% C.L.\ allowed
masses for the LW eigenstates $\tilde{t}_0^{(1,2)}$ and $\tilde b_0$
following from Fig.~\ref{fig:two}, as functions of the LW gauge mass
parameter $M_2$.}
\end{figure}

Figure~\ref{fig:two} shows our results for the choice
$m_{\tilde{h}}=750$~GeV; over the phenomenologically interesting range
$250$~GeV$<m_{\tilde{h}}<1$~TeV we find a remarkably weak dependence
of $S,T$ on $m_{\tilde{h}}$, and therefore opt to fix $m_{\tilde h}$
at an intermediate value.  The grid shows model predictions as
$M_2$ and $M_F$ are varied from $2$--$10$~TeV.  The 95\% C.L.\@ allowed 
region is based~\cite{gk} on an analysis by the LEP Electroweak Working 
Group~\cite{LEP}, but shifted to convenient Higgs and $t$ reference 
masses, 115~GeV and 170.9~GeV, respectively.  We find points with $M_2$ and $M_F$ 
both $\alt 5.2$ TeV just within the allowed region; for  $M_2=10$~TeV, $M_F$ can 
be as small as $\sim 4$~TeV. Fig.~\ref{fig:three} displays the 
smallest allowed masses for the LW eigenstates $\tilde{t}_0^{(1,2)}$ and $\tilde b_0$
following from Fig.~\ref{fig:two}.  For example, this figure indicates
that one of the $\tilde t_0$'s can be as light as $4$~TeV.  With LW gauge and
fermion states typically heavier than this, other low-energy constraints on the model, 
such as those from flavor-changing processes~\cite{Dulaney:2007dx,Krauss:2007bz}, are 
likely only to be relevant in the lighter LW Higgs sector.  A more complete investigation 
of flavor and electroweak constraints on the effective theory of interest will appear in 
Ref.~\cite{clnext}.

\section{Conclusions} \label{sec:concl}
We have considered the constraints from oblique electroweak parameters
in an extension to the SM that includes LW partners to the SU(2) gauge
bosons, the Higgs doublet, and the $t_{L,R}$ and $b_L$ quarks.  This
low-energy theory has the smallest particle content required to cancel
the largest contributions to the Higgs quadratic mass divergences,
rendering the effective theory natural up to $\sim 10$~TeV (similar to
Little Higgs models).  Above this scale one may uncover the remaining
particle content of the LWSM, or perhaps an even more exotic
ultraviolet completion.  This effective theory is meritorious because
its spectrum is simple and allows a more focused and complete study of
phenomenological constraints and collider signatures; the electroweak
analysis presented here is a necessary first step.  Our conclusion
that the LW partners in this effective theory can be kinematically accessible
at the LHC (as specified in Figs.~\ref{fig:two}--\ref{fig:three}) suggests 
a broad range of interesting phenomenological issues for further study.

\vspace{-0.5em}
\section*{Acknowledgments}
\vspace{-0.5em}
This work was supported by the NSF under Grant Nos.\ PHY-0456525 and
PHY-0757481 (CDC) and PHY-0456520 (RFL).  CDC thanks Arizona State
University and RFL thanks the Institute for Nuclear Theory for
hospitality during part of the time in which this work was performed.

\end{document}